\documentclass{aa}  

\usepackage{txfonts}
\usepackage{graphicx,lipsum,overpic}
\usepackage{subfigure}
\usepackage{amssymb,amsmath}

\newcommand{\ignore}[1]{}

\def\aap {{A\&A}}

\def\apjl {{ApJL}}

\def\apjs {{ApJS}}

\graphicspath{{../}}
\begin{document}

   \title{Magnetically induced termination of giant planet formation}


   \author{A.J. Cridland
          \inst{1}\thanks{cridland@strw.leidenuniv.nl}
          }

   \institute{
   $^1$Leiden Observatory, Leiden University, 2300 RA Leiden, the Netherlands
             }

   \date{Received \today}


  \abstract
  {
  Here a physical model for terminating giant planet formation is outlined and compared to other methods of late-stage giant planet formation. As has been pointed out before, gas accreting into a gap and onto the planet will encounter the planetary dynamo-generated magnetic field. The planetary magnetic field produces an effective cross section through which gas is accreted. Gas outside this cross section is recycled into the protoplanetary disk, hence only a fraction of mass that is accreted into the gap remains bound to the planet. This cross section inversely scales with the planetary mass, which naturally leads to stalled planetary growth late in the formation process. We show that this method naturally leads to Jupiter-mass planets and does not invoke any artificial truncation of gas accretion, as has been done in some previous population synthesis models. The mass accretion rate depends on the radius of the growing planet after the gap has opened, and we show that so-called hot-start planets tend to become more massive than cold-start planets. When this result is combined with population synthesis models, it might show observable signatures of cold-start versus hot-start planets in the exoplanet population.
}


    \maketitle
%
\section{ Introduction }\label{sec:intro}

While giant planet formation on the whole is a relatively slow process ($\sim$ few Myr) in the classical core-accretion scenario, its final stage of gas accretion occurs very quickly \citep{Pollack1996}. Once begun, a giant planet can accrete a Jupiter mass of gas on timescales as short as $\sim 10^4$ yr. With such short timescales, it is unlikely that the final stage of a protoplanetary disk - its evaporation - can be the sole terminator of planet formation. 

All large planets evolve through a phase in which they open a gap in their protoplanetary disk. When the planet is sufficiently massive, its gravitational influence can exceed the viscous forces in the disk, opening the gap \citep{Lin1986,Bryden1999}, and decoupling the planet from the surrounding gas disk. When this happens, the physical processes that govern planet migration (i.e., Type-II migration), and gas accretion change. It has been proposed that the termination of gas accretion might be linked to this decoupled phase of planet formation, but hydrodynamic simulations (see, e.g., \cite{Szul2014,Szul2016}) have shown that gas continues to flow into the gap of Jupiter-mass planets. Hence it is insufficient to simply open a gap to explain the final planet masses.

\cite{Morbidelli2014} discussed this terminal mass problem and argued that there is no obvious reason why the process of gas accretion should stop. In other words, we lack a complete physical picture to explain the final mass of Jupiter, Saturn, and the observed exoplanet population.

In this paper, we explore a simple physical picture for the termination of gas accretion on giant planets based on the recent work of \cite{Batygin2018}. This picture begins with the idea that gas accretion through a gap is predominately achieved through vertical circulation of gas from regions of the disk above the midplane \citep{Szul2014,Morbidelli2014}. As it falls toward the planet, the gas encounters a magnetic field that is being generated by a magnetic dynamo in the convective layer of the forming planet's interior. Within the magnetosphere (defined in Sect. 3) magnetic effects are stronger in the gas dynamics than gravity, and gas tends to flow along field lines rather than fall freely to the midplane. Hence gas can be funnelled away from the planet into the circumplanetary disk if it lands on the far side of the crest of the critical magnetic field line. When it is in the circumplanetary disk, the gas is recycled into the protoplanetary disk as it transfers angular momentum away from the planet. 

We show here how this simple physical picture can lead to stalled planetary growth. In Sect. \ref{sec:term} we outline the rate of gas flow into the gap in both a 1D and 3D picture. In Sect.  \ref{sec:maths} we summarize the physical model of \cite{Batygin2018} and derive a  gas accretion rate onto the planet. The planet formation model that is used to test the late-stage effect of this new accretion rate is outlined in Sect. \ref{sec:coreacc}. The results of the magnetically terminated gas accretion model are presented in Sect. \ref{sec:results}. We summarize and conclude in Sect. \ref{sec:con4}.

\section{Gas flow rate into a planetary gap}\label{sec:term}

In a 1D sense, the growth rate of a planet after a gap is opened is limited by the rate at which this gas can be delivered to the edge of the gap. It follows that the maximum gas accretion rate onto the planet is related to the global mass accretion rate through the disk (see, e.g., \cite{IL04a,Mordasini15}). Accretion disks tend to relax to a quasi-steady state \citep{PT99} with a mass accretion rate of \citep{Alb05} \begin{align}
\dot{M}_{1D} = 3\pi\nu\Sigma_g,
\label{eq:intro01}
\end{align}
where $\nu = \alpha H^2 \Omega$ is the gas viscosity in the standard $\alpha$-disk prescription \citep{SS73} (with $\alpha = 0.002$ in our disk model, discussed below), $\Sigma_g$ is the gas surface density, $H$ is the scale height of the gas, and $\Omega$ is the Keplerian frequency. 2D numerical simulations (i.e., \cite{Bryden1999}) have shown that material continues to flow through the gap, with some fraction of it being delivered to the planet. In this simple picture, the rate at which material is delivered can therefore be no quicker than in Eq. \ref{eq:intro01}. This provides a very natural if slow method of terminating gas accretion because the mass flux through a disk preferentially drops as the disk ages (see, e.g., \cite{Alessi16}). 

This picture becomes more complicated when the flow of material in 3D is considered. As pointed out by \cite{Morbidelli2014}, the typical 1D treatment of material flowing into the gap (i.e., Eq. \ref{eq:intro01}) ignores an important effect that arises from the vertical structure of the disk. In particular, if the disk is in hydrostatic equilibrium in the vertical direction, its gas density has the same radial dependency at all heights. However, the gravitational influence of the planet drops with height above the midplane, which allows material to cross the traditional gap boundary more easily from a height $z>0$ than it can from the midplane \citep{Morbidelli2014}.

When the gas enters the gap, it immediately loses its pressure support and rapidly falls to the midplane, where it is either accreted by the planet or filtered back into the disk. The gas that is filtered back into the disk increases the density along the disk midplane, which forces gas toward the disk surface to reinstate hydrostatic equilibrium \citep{Morbidelli2014}. This recycles material up to heights that can again flow into the gap.

\cite{Morbidelli2014} derived a faster rate of material delivery into the gap than in Eq. \ref{eq:intro01} because the circulation described above evolves on a dynamical timescale rather than on the viscous timescale. In this case, the mass accretion rate into the gap at a disk radius $r$ is\begin{align}
\dot{M}_{3D} = 2\pi r \left[\alpha (H/r)r\Omega\right] 4\Sigma_g,\nonumber
\end{align}
where the terms in the square bracket are the radial speed of the gas through the disk midplane, and the additional factor of 4 that precedes $\Sigma_g$ accounts for the inner and outer edges of the gap, and the top and bottom surface layers of the disk. 

We rewrite their expression to look more similar to Eq. \ref{eq:intro01}:\begin{align}
\dot{M}_{3D} = 8\pi \nu (r/H) \Sigma_g,
\label{eq:intro02}
\end{align}
using the definition of viscosity. The mass accretion rate depends on the inverse of the disk scale height ratio (H/r), which has a typical value of 5$\%$, hence $\dot{M}_{3D}/\dot{M}_{1D}\sim 50$.

As described above, not all of the material that is accreted into the gap ends up on the planet. In the 1D case, gas entering the gap is supported rotationally, entering into a circumplanetary disk. In this case the gas accretion rate onto the planet is limited by the accretion rate through the circumplanetary disk. This can in principle result in a different accretion rate onto the planet because the circumplanetary disk has a different physical structure (and hence different mass accretion rate) than the surrounding protoplanetary disk.

Owing to vertical circulation, the gas falls vertically toward the planet in the 3D case. Any gas that is not accreted onto the planet flows into a circumplanetary disk and is recycled into the protoplanetary disk. This physical picture contains two important differences to the 1D model: (1) the net direction of material flow through the circumplanetary disk (outward) is different here than in the 1D case (inward). Moreover, (2) while in the 1D case the accretion rate onto the planet is limited by the efficiency that gas can accrete through the circumplanetary disk, the limiting factor in the 3D case depends more on the accretion geometry.

Neither of these methods explicitly predicts the percentage of the material that is accreted onto the planet, and this is often modeled using some efficiency factor, such that\begin{align}
\dot{M}_{p} = \dot{M}\epsilon,
\label{eq:intro04}
\end{align}
where $\dot{M}$ is the mass accretion rate into the gap (either $\dot{M}_{1D}$ or $\dot{M}_{3D}$), and $\epsilon < 1$.  In the 1D case $\epsilon$ represents a parameterization of the viscosity in the circumplanetary disk, while in the 3D case, we propose that it is a geometric factor describing the fraction of gas that can accrete onto the planet.

Particular values of $\epsilon$ vary between different works: for example, \cite{Morbidelli2014} suggested $\epsilon = 0.5$ for their fiducial setup, while \cite{Bitsch2015} used $\epsilon = 0.8$ in their formation models. For both the 1D and 3D gas accretion models, planetary growth is eventually terminated because the surface density of the surrounding gas drops as the disk evolves. However, this rate can remain high enough in typical disks to overproduce very massive planets within 1 AU, unless their growth is artificially truncated.

Before exploring a more complicated model, we note that a current method of limiting the mass of a planet is by setting a maximum planetary mass, \begin{align}
M_{max}= f_{max} M_{gap},
\label{eq:intro03}
\end{align} 
above which the gas accretion is artificially stopped. This parameterization is related to the gap-opening mass ($M_{gap}$), which at least relates this method to the expectation that accretion termination is related to the planet opening a gap in the disk. This method has been used in both population synthesis models \citep{HP13,Alessi16b} and in end-to-end planet formation models \citep{Crid16a,Crid17} to simplify the final stages of planet accretion by ignoring the physical processes responsible for terminating planetary growth. In population synthesis models, f$_{max}$ is randomly selected from a range of 1-500, which, given that M$_{gap}\lesssim 10$ M$_{\oplus}$ , results in a population of giant planets with a mass range of $\sim$ 10 M$_\oplus$ - 15 M$_{J}$. While it is useful for statistical studies, f$_{max}$ has little connection to the underlying physics that result in the termination of gas accretion.

\section{Magnetically limited gas accretion}\label{sec:maths}

From 3D hydrodymamic simulations of gas accretion through a gap, \cite{Szul2014} showed that the bulk of gas that was accreted into the gap (90\%) did so by falling vertically from heights above the midplane of the disk. In the absence of planetary accretion, this gas formed a circumplanetary disk and then returned to the surrounding protoplanetary disk.

With the addition of a magnetic field, magnetic effects will alter the flow of the weakly
ionized gas within a characteristic radius ($R_t$, see below). As illustrated by \cite{Batygin2018}, within this radius, the flow is confined to the magnetic field lines. It enters into a force-free configuration (\textbf{v}$\times$\textbf{B}$\rightarrow$ 0) and is not falling freely. This means that when gas falls on the critical field line it flows onto the planet if it lands on the planet side of the line apex. Otherwise it is deflected away from the planet to join the circumplanetary disk.

Given a flow of material into the gap, $\dot{M}$, the characteristic radius within which magnetic effects dominate the dynamics of the gas is \citep{Batygin2018}\begin{align}
R_t &= \left(\frac{\pi^2}{2\mu_0}\frac{\mathcal{M}^4}{GM_p\dot{M}^2}\right)^{1/7}\nonumber\\
R_t &= R_0 \left(\frac{M_p}{M_\oplus}\right)^{-1/7}\left(\frac{\dot{M}}{M_\oplus/yr}\right)^{-2/7},
\label{eq:math01}
\end{align}
where $R_0 = (\pi^2/2\mu_0 ~\mathcal{M}^4/GM_\oplus^2/yr)^{1/7}$ has units of length, and $\mathcal{M} = B R_p^3$ is the magnetic moment of an (assumed) magnetic dipole. We follow \cite{Batygin2018} in setting the strength of the magnetic field as $B=500~Gauss,$ which is about two orders of magnitude stronger than the magnetic field of Jupiter today, but weaker than the typical field strength derived for $\sim 1000$ K brown dwarfs. We select a radius of the young planet $R_p = 2R_J$ as it is a nominal size for young, self-luminous planets with masses $\gtrsim 100$ M$_\oplus$ reported by \cite{Mordasini15}. We assume that both of these quantities remain constant throughout the latter stages of planet formation (see below).

Neither of these physical quantities is well constrained by observations, and their evolution during the phase of rapid gas accretion is not known a priori{\it } in our model. As argued in \cite{Batygin2018}, a sub-kilo Gauss magnetic field is consistent with our understanding of magnetic dynamo theory for young, self-luminous, Jupiter-mass planets. Furthermore, as illustrated by \cite{Christensen2009}, a sub-kilo Gauss magnetic field is a good interpolation for what we understand of the geo and solar dynamo. 

The source of the magnetic field are the convective cells that transport heat through the internal structure of the gas giant \citep{Christensen2009}. If the field is in equipartition with the kinetic energy of the convective cell, the strength of the magnetic field should roughly scale as $B\propto \sqrt{\rho v^2_{conv}}$ \citep{Christensen2009,Batygin2018}, where $\rho$ is the average density of the convecting region, and $v_{conv}$ is the convection speed. This speed is related to the energy transport rate through convection, and hence on the internal energy flux of the planet. 

While the total luminosity of the planet greatly increases during gas accretion \citep{Mordasini13}, this energy largely comes from the release of gravitational energy by the accreting gaseous envelope. Some of this energy is absorbed by the planet, and hence the convection rate can in principle change as the planet grows. The amount of energy that is absorbed by the planet is the distinguishing factor for two formation scenarios: the so-called hot- and cold-start models. In the hot-start model, some of the accretion luminosity is absorbed by the planet and increases its internal temperature. Conversely, in the cold-start model, the majority of the accretion luminosity is radiated away. If the magnetic field remains in equipartition with the convection cells, these two scenarios may generate different strengths and evolutions of the magnetic field. However, we keep the magnetic field constant for simplicity.

\cite{Mordasini13} computed a self-consistent model for the internal structure of accreting gas giants. Of planetary properties that he computed, he showed the evolution of the planet's outer radius in his Fig. 1. When a planet opens its gap and detaches from the surrounding disk, it rapidly reduces its radius from approximately the Hill radius ($R_H$, see below) to a few Jupiter radii (see Fig. 1 in \cite{Mordasini13}), depending on the entropy of the planet when the gap is opened. For a hot-start planet (high entropy), the outer radius evolves to between 3-4 R$_J$ for the remainder of its gas accretion. For a cold-start planet (low entropy), its radius evolves to $\sim 1.5$ R$_J$ after the gap is opened.  Our choice of 2 R$_J$ therefore represents an average value of these two scenarios. After the gap opens, the planetary radius shows little evolution \citep{Mordasini13}, which means that keeping it constant during the latter stages of planet formation seems reasonable.
Because of its contrasting formation scenarios, we also compute the formation of a cold-start planet with $R = 1.5$ R$_J$ and a hot-start planet with $R = 3$ R$_J$. 

Within a distance of $R_t$ from the planet, the gas dynamics is dominated by magnetic effects, confining the gas to the field lines rather than allowing a free fall onto the midplane. This confinement requires low gas resistivity, such that the magnetic Reynolds number ($Rm$) is larger than one. This can most easily be attained at temperatures $>$ 1000 K when alkali metals are thermally ionized. Such gas temperatures are readily produced around young forming planets in hydrodynamic simulations (see, e.g., \cite{Szul2016}). Additionally, when a gap is opened in the disk, the gas can become more susceptible to X-ray ionization, as argued by \cite{Tan2013}, further raising the Reynolds number. Here we generally assume that the incoming gas is suffienctly ionized such that $Rm>>1$ and it remains confined to the magnetic fields.

As the gas falls on the critical field line, it flows onto the planet if it lands planet-ward of the field line crest, otherwise, it flows along the field line onto the circumplanetary disk. When it is in the disk, the gas flows back into the surrounding protoplanetary disk as it carries angular momentum away from the planet \citep{Batygin2018}. This naturally sets a vertical cross section that will limit the mass accretion rate onto the planet, which for a dipole field is roughly $A_{mag}\sim \pi R_t^2/3^{3/4}$ \citep{Batygin2018}. Comparing this cross section to the total accreting area around the planet allows us to derive an accretion efficiency ($\epsilon$).

We assume that the total accretion area is the cross section of the circumplanetary disk. \cite{Szul2014} ran numerical simulations of gas accretion onto giant planets through a gap. They reported radii of the resulting circumplanetary disks that ranged between $\sim 0.28 - 0.75$ Hill radii ($R_{H} = a(M_p/3M_*)^{1/3}$) depending on their (numerical) viscosity. The edge of the circumplanetary disk connects to the protoplanetary disk at the gap edge. If we assume that the material accreting into the gap falls homegeneously across the planet-cirumplanetary disk system, then the accretion efficiency ($\epsilon$) would be the ratio of $A_{mag}$ and the cross section of the circumplanetary disk: $A_{grav} = \pi R_H^2/4,$  where we chose a nominal circumplanetary disk radius of 0.5$R_{H}$. This means that\begin{align}
\epsilon &= A_{mag} / A_{grav}\nonumber\\
&= \frac{4 R_t^2}{3^{3/4} R_H^2}\nonumber\\
&= \frac{4}{3^{3/4}}\left(\frac{R_0}{R_H}\right)^2\left(\frac{M_p}{M_\oplus}\right)^{-2/7}\left(\frac{\dot{M}}{M_\oplus/yr}\right)^{-4/7}.
\label{eq:math02}
\end{align}

Noting again that $\dot{M}$ is the mass accretion rate into the gap, we combine Eqs. \ref{eq:intro04} and \ref{eq:math02}: \begin{align}
\frac{\dot{M}_{p,Mag}}{M_\oplus/yr} = \frac{4}{3^{3/4}}\left(\frac{R_0}{R_H}\right)^2 \left(\frac{M_p}{M_\oplus}\right)^{-2/7}\left(\frac{\dot{M}}{M_\oplus/yr}\right)^{3/7}.
\label{eq:math03}
\end{align}
This shows that the gas accretion rate is stifled as the planet grows. There are two mass-dependent terms in Eq. \ref{eq:math03}. We recall that $R_H\propto M_p^{1/3}$, and the overall mass dependence of the accretion rate is $\dot{M}_{p,Mag}\propto M^{-20/21}$.

We can interpret eq. \ref{eq:math03} in two ways. First, as the planetary mass grows, the region where the magnetic force is relevant to the gas dynamics shrinks, and so too does the planet accretion cross section. Second, the physical size of the gap grows with the planetary mass. If the mass is accreted homogeneously across the total disk cross section, the quantity of mass available within $R_t$ also shrinks naturally.

The general picture is as follows: the gas that falls toward the planet outside of R$_t$ has too much angular momentum relative to the planet to accrete directly, and falls onto the circumplanetary disk. As argued by \cite{Batygin2018}, the disk carries rotational angular momentum away from the planet, transporting material back to the protoplanetary disk (as required by \cite{Morbidelli2014}). Gas that falls within R$_t$ is affected by the magnetic field, and will accrete onto the planet along the magnetic field lines as long as it lands within the cross section defined by A$_{mag}$. Otherwise, it flows along the field line to the circumplanetary disk, away from the planet. As the planet grows, its gravitational influence grows. This decreases the cross section within which gas can continue to accrete onto the planet; this eventually stifles further growth.

\section{ Core accretion model }\label{sec:coreacc}

To test the effects of this new gas accretion terminator, we modified the final stages of the planet formation model outlined in \cite{Alessi16} and \cite{Crid17}. These models assumed the standard planetesimal accretion scheme \citep{KI02,IL04a} with a planet-trapping model of planet migration \citep{HP11} to build the initial $\sim 10$ M$_\oplus$ planetary core. When the growing core is sufficiently massive, it begins a phase of gas accretion that is first limited by the Kelvin-Helmholtz time \citep{Ikoma2000}: \begin{align}
t_{KH} = 10^c yr\left(\frac{M_p}{M_\oplus}\right)^{-d},
\label{eq:meth01}
\end{align}
such that\begin{align}
\dot{M}_{p,KH} = \frac{M_p}{t_{KH}},
\label{eq:meth02}
\end{align}
where the parameters $c$ and $d$ depend on the opacity of the accreting planetary envelope $\kappa_{env}$ \citep{Ikoma2000,Miguel2008,Mordasini14,Alessi16b}. Recently, \cite{Alessi16b} tested fitted functions for both $c(\kappa_{env})$ and $d(\kappa_{env})$ based on the work of \cite{Mordasini14} using population synthesis models and found that a preferred value of $\kappa_{env} = 0.001$ cm$^2$g$^{-1}$ best reproduced the population of known exoplanets. Using their preferred values, we chose $c = 7.7$ and $d=2.0$.

Eventually, the growing planet becomes sufficiently massive for its gravitational influence to exceed the fluid dynamics that otherwise govern the structure of the disk. This is characterized by either the gravitational torque of the planet on the surrounding material that exceeds the disk viscosity torque, or the gravitational influence of the planet ($R_H$) that exceeds the typical length scale of the gas pressure ($H$). When one of these requirements is met, we say that the planet has opened a gap. This has a characteristic gap-opening mass of \citep{MP06}\begin{align}
M_{gap} = M_* {\rm min}\left(3h^3,\sqrt{40\alpha h^5}\right),
\label{eq:meth03}
\end{align}
where $h=H/r$ is the disk scale height ratio. The first term represents the mass required for $R_H = H$, while the second term represents the mass required for the gravitational torque of the planet to exceed the disk viscous torques.

When the gap is opened, the magnetically limiting gas accretion term from Eq. \ref{eq:math03} can begin to affect the flow rate of gas onto the planet. However, if the rate exceeds $\dot{M}_{p,KH}$ , then the gas cannot lose its excess gravitational potential energy fast enough to fully accrete onto the planet. The actual accretion rate is therefore the lower of the two rates,\begin{align}
\dot{M}_{p} = {\rm min}\left(\dot{M}_{p,KH},\dot{M}_{p,Mag}\right).
\end{align}

The method that builds the original solid core onto which gas is accreted remains a debated topic. The classical core accretion model (now known as planetesimal accretion, e.g., \cite{Pollack1996}, \cite{KI02}, and \cite{IL04a}) posits that the protoplanetary core is built from the successive collisions of 10-100 km planetesimals onto a $\sim 0.01-0.1$ M$_\oplus$ embryo. A potentially catastrophic issue with this model is that its typical timescale ($\sim 10^5$ yr) is about an order of magnitude longer than the Type-I migration rate for planets with masses of $\sim$ M$_\oplus$ \citep{Masset06}. Planets should therefore be lost through disk-planet interactions before they could accrete to masses high enough to accrete gas (few M$_\oplus$). 

Two models have been suggested to remedy the Type-I migration problem: planet trapping, and pebble accretion. The former slows the Type-I migration rate through disk inhomogeneities, which leads to positions in the disk where the net torque on the planet reaches zero; this is often called planet traps (or convergence zones, see the review by \cite{Pudritz2018}). The latter remedy assumes that solid accretion is dominated by centimeter-size particles that accrete at a much faster rate because they are less susceptible to gravitational scattering than planetesimals (see the review by \cite{Morbidelli2018}).

While the details of core accretion are less important to our current discussion (in as much as we require a core to form and accrete gas), it is worth noting that the rate of solid core accretion could position the protoplanet in such a way that it does not accrete an excessive amount of gas before the gas disk evaporates. In this way, we can view super-Earth planets as failed Jupiter cores that either did not have enough time to accrete a significant gaseous envelope, or grew in an excessively low density environment to undergo runaway gas accretion.

The planets undergoing magnetically terminated gas accretion all have opened a gap (by construction) and hence are not subject to Type-I migration. Instead, they enter into the second stage of planet migration: Type-II. In Type-II migration, the planet acts as an intermediary for the global angular momentum transport in the disk, and shrinks its orbital radius on the viscous timescale $t_\nu\sim \nu/r_p^2$, such that\begin{align}
\frac{1}{r_p}\frac{d r_p}{dt} = -\frac{\nu}{r_p^2}.
\label{eq:math04}
\end{align}
This rate persists until the planetary mass exceeds the mass of the disk gas inward of the planet $M_{crit}\sim \pi\Sigma r_p^2$. In this regime the migration rate is said to be limited by the planet's own inertia,\begin{align}
\frac{1}{r_p}\frac{d r_p}{dt} = -\frac{\nu}{r_p^2}\left(1 + M_p/M_{crit}\right)^{-1}.
\end{align}

We use the standard planetesimal accretion scenario below for a planetary embryo trapped at the water ice line (see \cite{Crid16a} and \cite{Alessi16b} for technical details).

\section{Results}\label{sec:results}

To show the effect of the magnetically terminated gas (MTG, eq. \ref{eq:math03}) accretion rate, we compare it to the 1D (Eq. \ref{eq:intro01}) and 3D (Eq. \ref{eq:intro02}) gap accretion rates as discussed in Sect.  \ref{sec:term}. Following \cite{Bitsch2015} and \cite{Morbidelli2014}, we assumed that the efficiency ($\epsilon$) is 0.8 and 0.5 for the 1D and 3D accretion rates, respectively. We ran our planet formation model over 7 Myr of evolution, which would constitute a very old disk, at least one standard deviation outside of the mean age of protoplanetary disks (see \cite{Alessi16b} for a theoretical distribution of disk ages). Where needed, we imposed a maximum mass using $f_{max}=100$ for all models other than MTG.

We built planets in the simple evolving-disk model reported by \cite{Eistrup2017}. The model fits a radial power law to the temperature and surface density profiles  computed by \cite{Alibert2013}. As the model evolves, the gas cools and the surface density drops, as expected from a viscously accreting protoplanetary disk. In this way, this model reflects the way in which the decreasing mass accreting rate through the disk can terminate the growth of giant planets. In this disk model, the mass accretion rate as computed with Eq. \ref{eq:intro01} is only a few $\times 10^{-10}$ M$_\odot$/yr, which is at the low end of the typical mass accretion rates for disks: $10^{-7}-10^{-10}$ M$_\odot$/yr. To account for the possibility that the mass accretion rate through the disk could be higher, we also tested a model where the 1D gap accretion rate was increased by an order of magnitude. Below we call this model the enhanced 1D gap model.

For the gap accretion rate ($\dot{M}$) in eq. \ref{eq:math03} we use the 3D accretion rate from eq. \ref{eq:intro02} because it represents (as we will see) the highest accretion rate we model. This will illustrate the utility of MTG as a method of limiting gas accretion. 

\begin{figure*}
\centering
\subfigure[Temporal evolution of gas accretion rate]{
\includegraphics[width=0.5\textwidth]{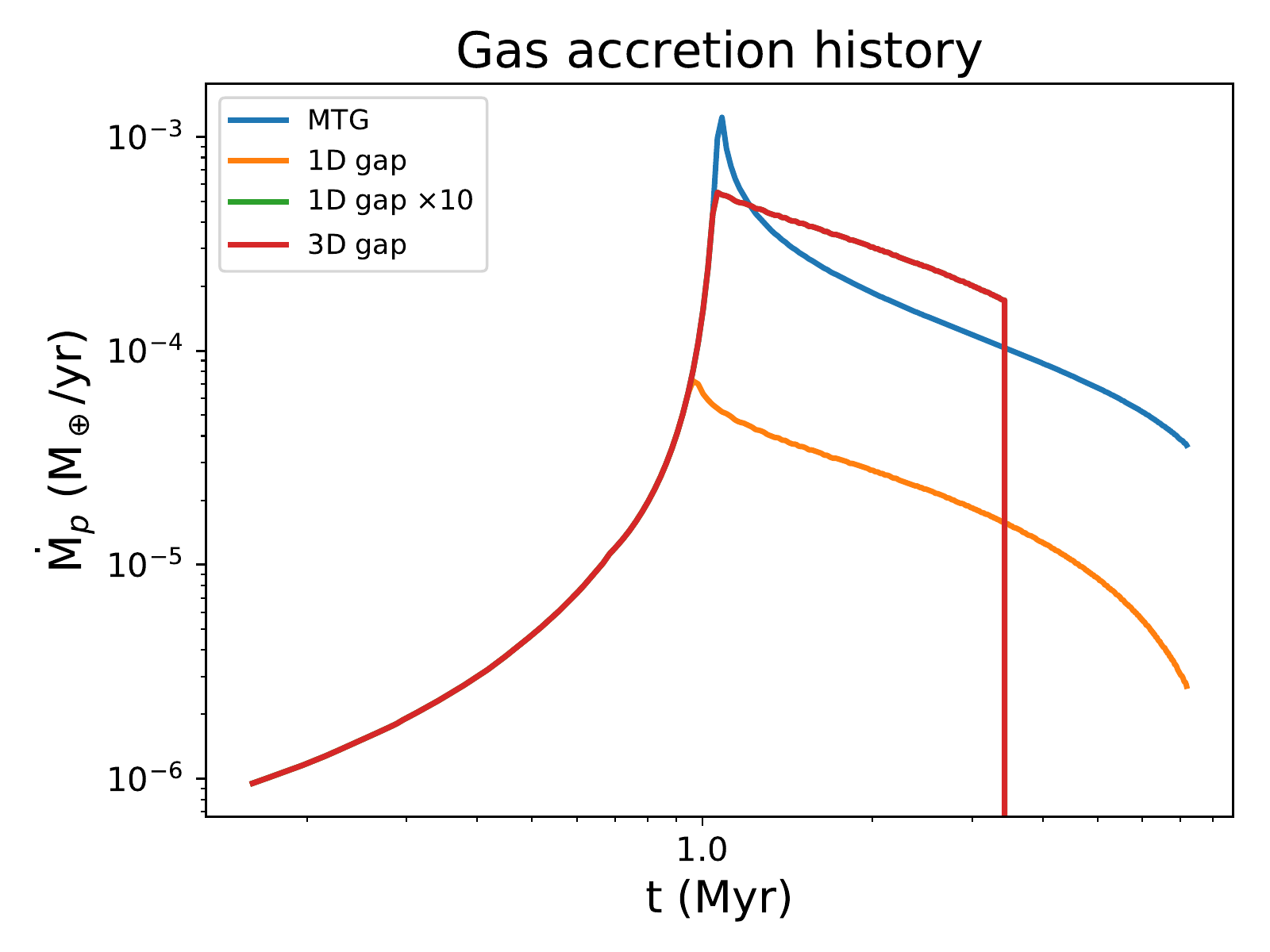}
\label{fig:res01a}
}
\subfigure[Temporal evolution of planet mass]{
\includegraphics[width=0.5\textwidth]{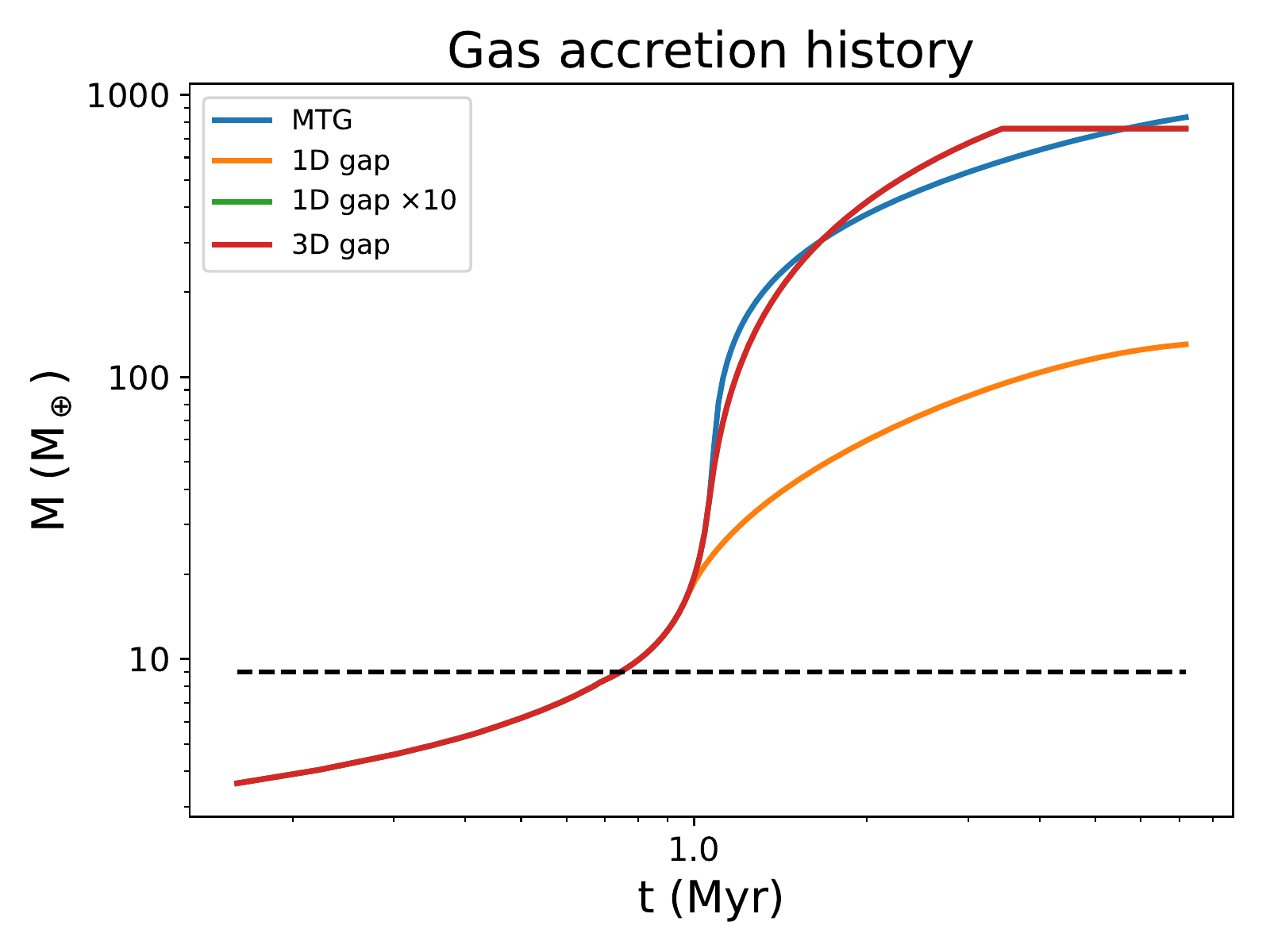}
\label{fig:res01c}
}
\subfigure[Mass doubling time: $t_{acc}= M/\dot{M}$]{
\includegraphics[width=0.5\textwidth]{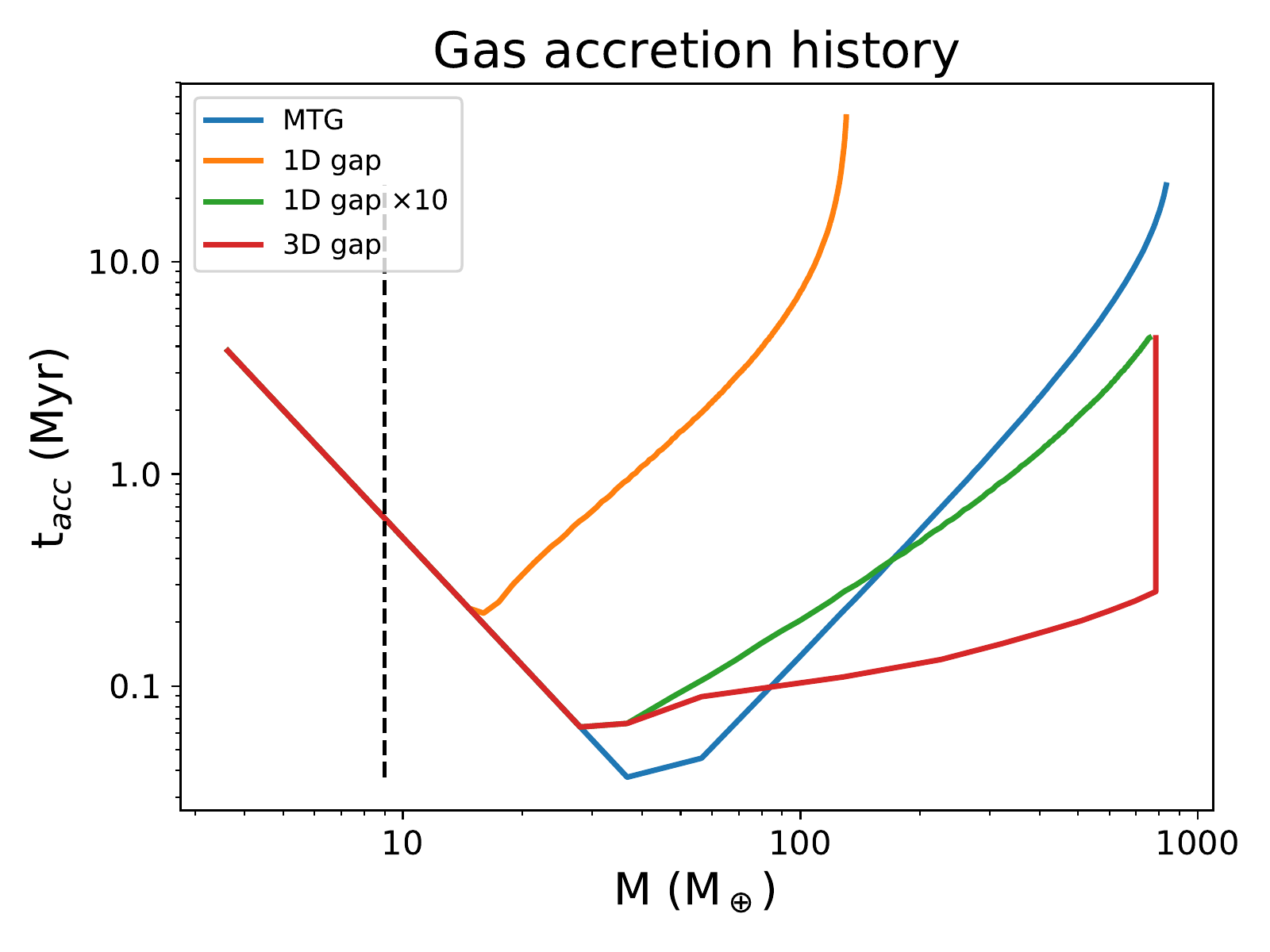}
\label{fig:res01d}
}
\subfigure[Mass dependence on the gas accretion rate]{
\includegraphics[width=0.5\textwidth]{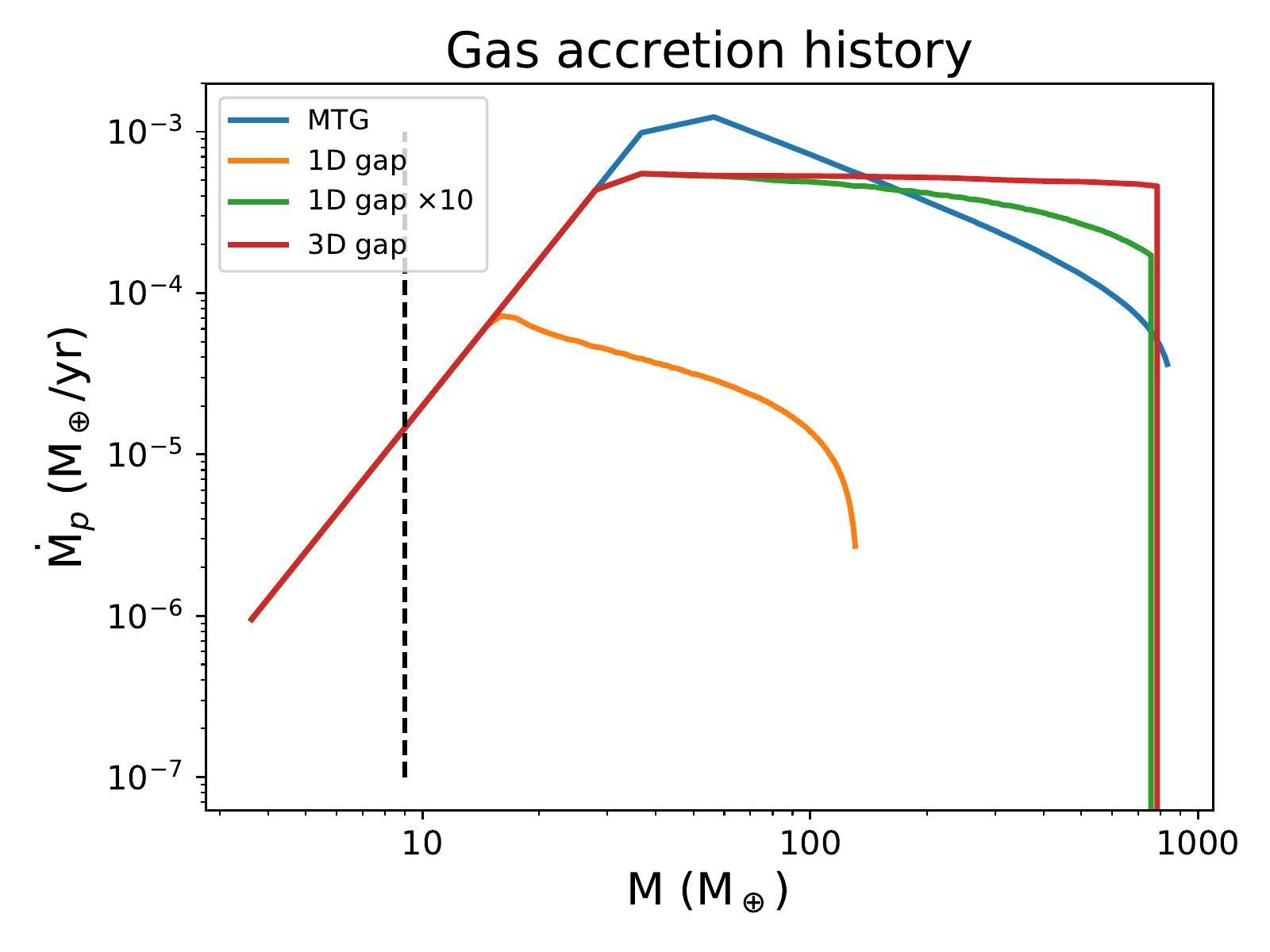}
\label{fig:res01b}
}
\caption{Evolution of the gas accretion history. For the MTG (Eq. \ref{eq:math03}), 1D disk accretion into the gap (eq. \ref{eq:intro01}), and the 3D gap accretion as prescribed by \citet{Morbidelli2014} (Eq. \ref{eq:intro02}). We increased the rate of the 1D disk accretion by a an order of magnitude to account for the possibility that gas accretion into the gap is increased by the generation of gravitational instabilities near the gap edge. For the 1D and 3D accretion rates, we assumed static efficiencies of 0.8 and 0.5, respectively. In each of the subfigures, the black dashed line denotes the point where the planet opens a gap.}
\label{fig:res01}
\end{figure*}

\subsection{Gas accretion history}

In Figure \ref{fig:res01} we show different representations of the formation histories of planets that grow with the four different accretion models: the MTG model, the 1D gap model, the enhanced 1D gap model (1D $\times 10$), and the 3D gap model. Each planet begins at the same point in the disk at the same point in the disk history. Each planet undergoes the same early evolution, and hence they all open a gap in their disk at the same point in time, denoted by the black dashed line. The main differences between the planets are at which point they break away from the unstable gas accretion phase (shortly after gap opening), and the rate at which material is accreted at later times. Below we summarize some key points.

In Figures \ref{fig:res01a} and \ref{fig:res01c} we show the temporal evolution of the planetary mass accretion rate and planet mass, respectively. In both the enhanced 1D and 3D accretion models, the planet reaches a maximum mass (Eq. \ref{eq:intro03}) that was set at $f_{max} = 100$. They are nearly indistinguishable in these figures, as they follow similar accretion histories. This is important to note because in population synthesis models that either have high disk masses or have planets that grow early in the disk lifetime, the choice of $f_{max}$ becomes the key parameter in determining the final mass of the planet. Conversely, in the 1D gap accretion model, the accretion rate is so low that the planet never approaches the maximum mass, therefore planets forming in lower mass disks or later in the disk lifetime are less sensitive to the choice of $f_{max}$.

The planet growing with the MTG accretion rate coincidentally grew to a mass that was similar to the maximum mass set by $f_{max}$. However, it took the full 7 Myr of growth to do so. This implies that the disk lifetime becomes more important in determining the final mass of the planet than in the accretion models that were limited by $f_{max}$. 

In Figure \ref{fig:res01d} we show an accretion timescale that effectively is a mass-doubling time, defined as the ratio between the current mass of a planet and its current mass accretion rate. It becomes clear why the MTG accretion model is an effective way of limiting gas accretion while still allowing for planetary growth up to Jupiter mass planets: the mass doubling time for a  $\sim 100$ M$_\oplus$ object is $\sim 0.2$ Myr, but this rapidly increases as the planet approaches a Jupiter mass. The enhanced 1D and 3D models both have mass doubling times of $\sim 0.1$ Myr at the same mass, but increase more slowly than the MTG model, which explains their need for setting a maximum mass. The 1D model has a mass doubling time closer to $\sim 10$ Myr at $\sim$ 100 M$_\oplus$ , resulting in its lower final mass.

In Figure \ref{fig:res01b} we show the mass dependence on the planetary gas accretion rate for each of the models. Here we see the important feature of the MTG accretion model: as the planet grows, it becomes harder to gather more material. This is because the planetary magnetic field diverts gas into the circumplanetary disk that otherwise would have been accreted onto the planet, if that gas reaches past the crest of the critical magnetic field line. Differing greatly from the other three mass accretion models that only limited the gas flow based on the amount of available material flowing toward the planet, MTG is self-regulating. 

\subsection{Planet formation history}

\begin{figure*}
\begin{minipage}{0.49\textwidth}
\centering
\includegraphics[width=\textwidth]{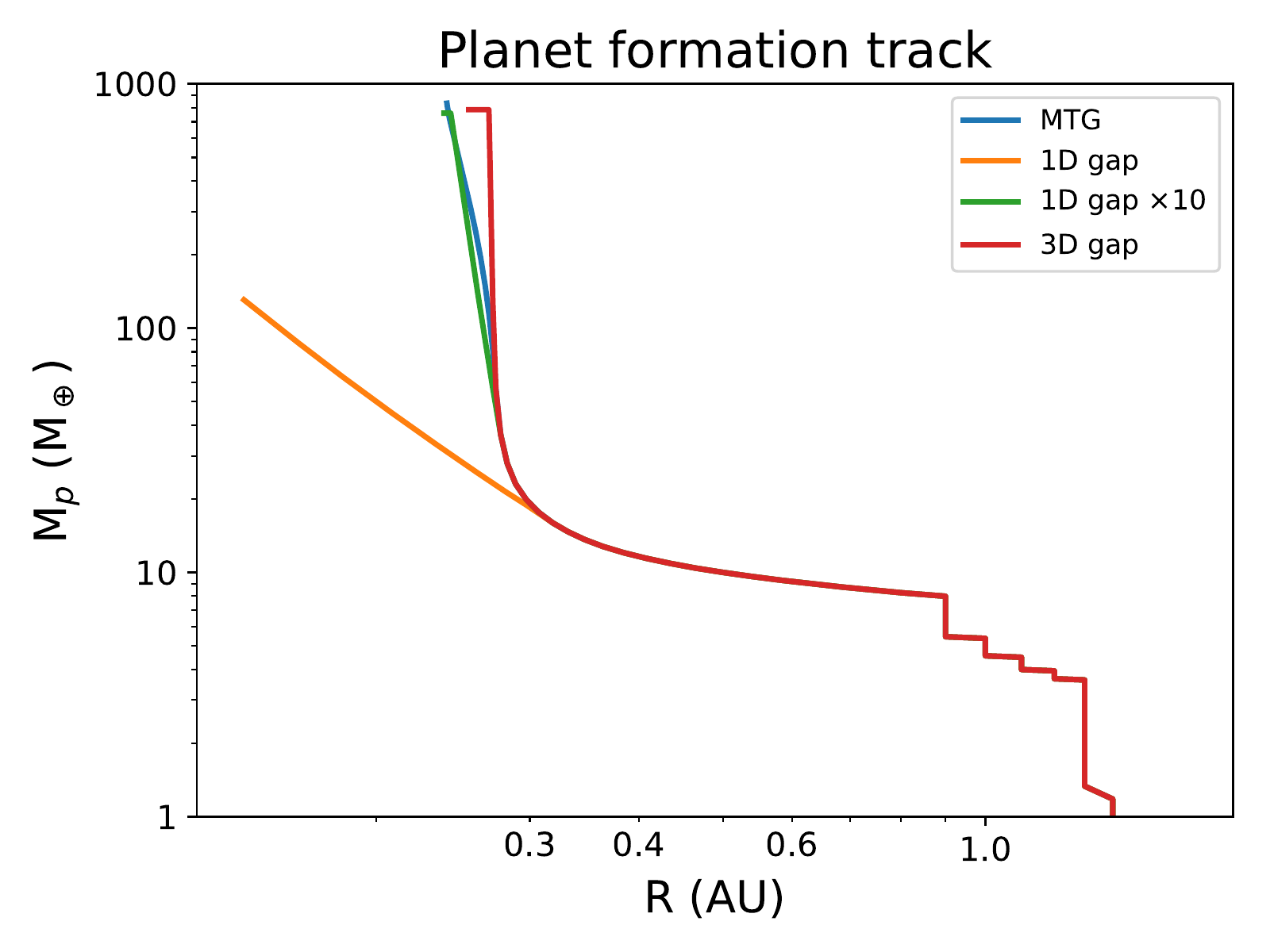}
\caption{Planet formation tracks for the four accretion models. The reason for the step-like evolution of the low-mass planet is that the radial evolution of the planet was dictated by planet-trapping at the water ice line, and therefore depended on the resolution of our disk model. The water ice line sometimes does not move between time steps, hence neither does the planet.}
\label{fig:res02}
\end{minipage}
\begin{minipage}{0.02\textwidth}
\hfill
\end{minipage}
\begin{minipage}{0.49\textwidth}
\centering
\includegraphics[width=\textwidth]{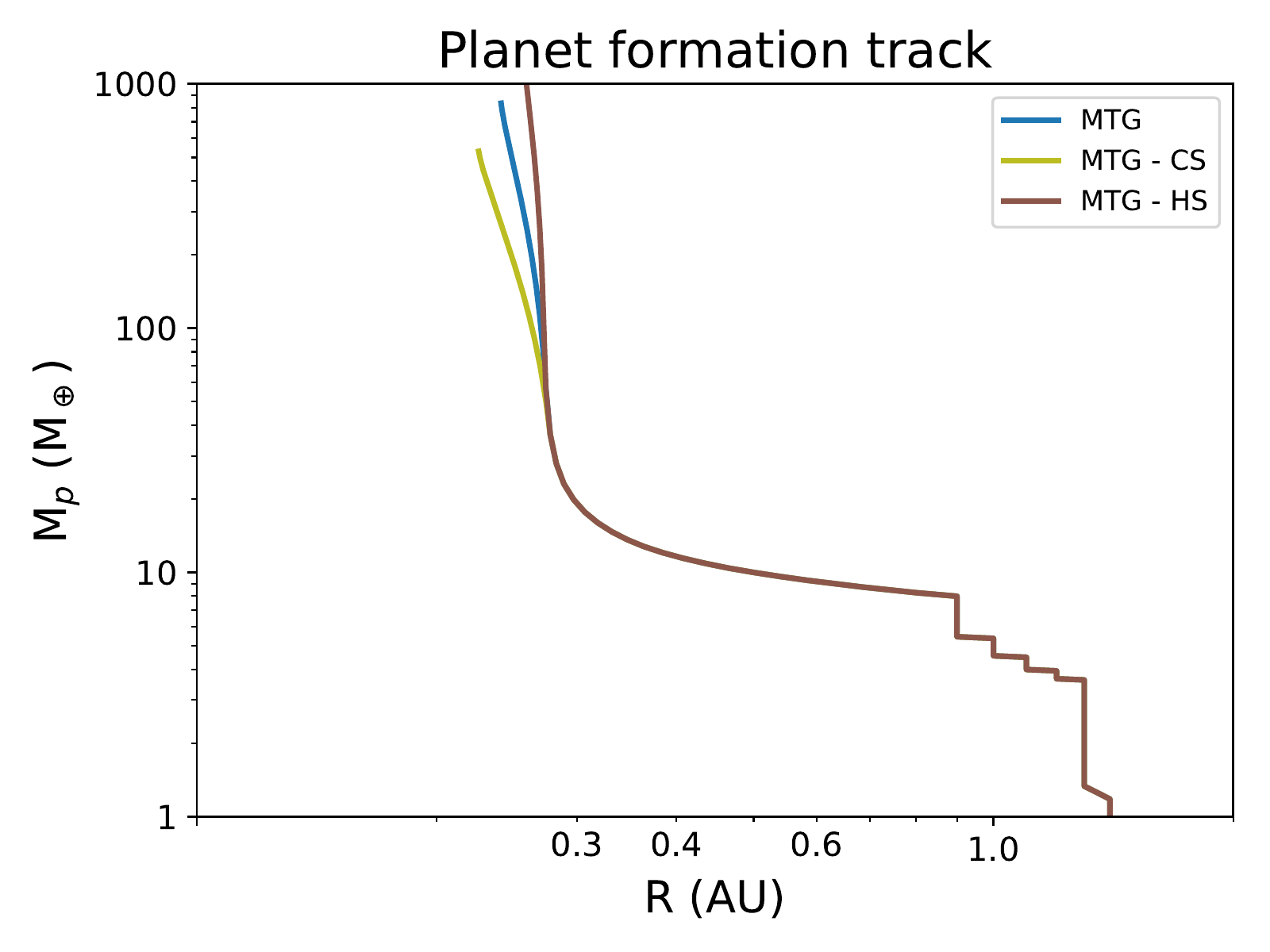}
\caption{Same as in Figure \ref{fig:res02}, but comparing the fiducial MTG accretion model to the HS and CS variants. By varying the size of the accreting planet, we change both the final mass and the radius to which it grows. When combined with the assumed age of the disk, we can similarly tune the final mass of the planet as with $f_{max}$.}
\label{fig:res03}
\end{minipage}
\end{figure*}

A common way of studying the formation history of synthetic planets is through its planet formation track. This track maps the planetary evolution through the commonly used mass-semi-major axis diagram.

In Figure \ref{fig:res02} we show the formation tracks for the four gas accretion models. While we started the process of solid accretion with an embryo of 0.01 M$_\oplus$, we only plot the formation track for masses $>$ 1 M$_\oplus$. The initial step-like growth between $\sim 1-10$ M$_\oplus$ is related to our prescribed location of the water ice line (in which the planet is trapped). The location of the water ice line is defined as the radius farthest inward where water ice becomes more abundant than gaseous water. Its particular location is selected as the disk radius with a gas temperature closest to 170 K, and is therefore limited by the spacial resolution of our temperature profiles. In many instances, the water ice line and hence the planet does not move between time steps. Ultimately, this step-like evolution affects the current discussion very little.

The formation tracks are identical until the planet opens a gap at $\sim 0.35$ AU where the migration scheme changes to the classic Type-II migration, and the gas accretion scheme changes to the scenarios outlined above. At this point, since the accretion timescales are all different (see Figure \ref{fig:res01d}), the tracks deviate from each other. The slower accretion rate (1D gap) evolves more in the horizontal direction than the faster accretors (MTG, 1D gap $\times 10$, and 3D gap) because its accretion timescale is equal to the viscous timescale, which dictates the rate of inward migration in the Type-II regime. When the fast accretors open a gap, their evolution is predominately vertical since their growth rates are faster than the Type-II migration rate. We note that Type-II migration eventually becomes very inefficient as the mass of the planet grows above the total gas mass within its orbital radius. This means that even when the growth timescale of MTG becomes long (we recall Figure \ref{fig:res01c}), its Type-II migration is too slow to change the orbital radius of the planet.

The three fast accretors all complete their evolution with similar planetary parameters (orbital radius and mass), but as noted earlier, a key difference is that the MTG accretion model does not require an artificial termination of its gas accretion.

\subsection{Hot versus cold starts}

The size of the planet when it enters the gap-opening phase of gas accretion is unknown in our model, and has the strongest effect on the resulting formation history of the planet because $\dot{M}_p\propto R_p^{24/7}$. According to \cite{Mordasini13}, the range of these sizes is between 1.5 and $\sim 3$ R$_J$ depending on whether the planet formed with a so-called cold (low entropy) or hot start (high entropy), respectively. Here we pick $R_p=1.5   \text{ and } 3$ R$_J$ and repeated the formation.

In Figure \ref{fig:res03} we compare the planet formation tracks for the fiducial model ($R_p=2$ R$_J$) to the cold (CS) and hot (HS) start variants. The main differences between the three lines are the turn-off points of the track, where MTG accretion begins to affect the planetary growth and also the final planet mass. Larger planets have larger magnetic moments, hence they start with a larger magnetic cross section and more efficient accretion (i.e., via Eq. \ref{eq:math02}). 

Its clear that in MTG accretion the size of the planet after it opens a gap has a strong influence on the late stages of gas accretion, hence it acts like $f_{max}$ from population synthesis models. However, unlike $f_{max}$, the planetary size depends sensitively on the processes that govern the evolution of its internal energy. To compare the two methods, we computed an {\it \textup{effective}} $f_{max,e} = M_{final}/M_{gap}$. For the cold-start, fiducial, and hot-start models, we find that $f_{max,e} \sim 160,244,\text{and }465$, which are on the high side of the original range of 1-500 that was used in population synthesis models. Of course these calculations were run on an older-than-average disk, and younger disks will result in lower values of $f_{max,e}$. A tantalizing question arises from this: is there an observable bias in the planet population toward a hot or cold start? Answering this question is beyond the scope of this paper, however, and requires a population synthesis model that includes MTG accretion.

\section{ Conclusions }\label{sec:con4}

Here we have outlined a simple physical model for the termination of gas accretion on a giant planet. This MTG accretion model is based on the work of \cite{Batygin2018}, who argued that the shrinking magnetosphere (defined by $R_t$, Eq. \ref{eq:math01}) would naturally limit the available cross section through which gas can accrete onto the planet. Material outside the crest of the critical magnetic field line falls past the planet onto the circumplanetary disk and is filtered back into the surrounding protoplanetary disk. This circulation was described by \cite{Morbidelli2018} and generally results in accretion rates that are between one and two orders of magnitude higher than would be predicted by simple 1D accretion of material through the protoplanetary disk.

When we take the ratio of the magnetic cross section with the cross section of the circumplanetary disk, and assume that material accretes vertically and homogeneously across the gap, this results in an efficiency factor $\epsilon$ that scales inversely with the mass of the growing planet. This naturally leads to a truncation in planetary growth because the magnetic cross section decreases as the planet grows. Additionally, by construction, the efficiency decreases with the planet mass because the gap size increases with planetary mass (through the Hill radius). Homogeneous accretion across the gap therefore implies that less material is available within the magnetic cross section even if R$_t$ were constant.

We are forced to assume (for simplicity in the model) that the magnetic field strength generated by the planetary dynamo as well as the size of the planet remains constant during the final stages of gas accretion, after the gap has opened. There is evidence suggesting that the planet size remains constant after gap opening based on formation calculations that include the internal structure of the planet and its evolution (i.e., \cite{Mordasini13}). The dynamo process responsible for the generation of the magnetic field is not constrained by observations, nor is it well understood theoretically. However, if is in equipartition with the convective flux in the interior of the planet, that would suggest that it remains largely unchanged while the planet accretes its outer envelope.

Because the final size of the planet after gap opening is unknown and depends on the internal processes during formation, we varied the planetary size $R_p$ and tested the resulting formation history of the planet. The range of $R_p$ was determined by the choice of a hot- or cold-start model, which depends on whether the planet has maintained high or low entropy, respectively, during its initial accretion. The choice of $R_p$ is effectively similar to the choice of $f_{max}$ in population synthesis models, but is more closely tied to physical processes.

In this framework, the disk lifetime becomes an incredibly important parameter to the population of giant planets. Moving forward, full population synthesis models must be used to test which range of planetary masses can be achieved with this method. The key questions are the maximum mass that is attainable using MTG accretion as the limiting model, and whether it can reproduce the range of planetary masses that we observed given a typical spread in disk lifetimes.

\begin{acknowledgements}

I gratefully acknowledge the anonymous referee for their insightful comments that greatly improved the manuscript. AJC acknowledges financial support from the European Union A-ERC grant 291141 CHEMPLAN, supervised by Ewine F. van Dishoeck. Thanks to Ralph Pudritz and Matthew Alessi for fruitful discussions during the early stage of this project.

\end{acknowledgements}

\bibliographystyle{aa} 
\bibliography{mybib.bib} 

\end{document}